\documentclass[twocolumn]{aastex62}

\usepackage{comment}
\usepackage{soul}

\newcommand\mdot{M$_{\odot}$}

\newcommand\mosfit{\href{https://mosfit.readthedocs.io/en/latest/}{MOSFiT}}
\newcommand\OSC{\href{https://sne.space/}{OSC}}
\newcommand\webpimms{\href{https://heasarc.gsfc.nasa.gov/Tools/multimissiontools.html}{WebPIMMS}}

\graphicspath{{./}{figures/}}

\received{---}
\revised{---}
\accepted{---}

\submitjournal{---}

\shorttitle{SN 2015bn: 2-year deep X-ray}
\shortauthors{Bhirombhakdi et al.}

\begin{document}

\title{Where is the engine hiding its missing energy? Constraints from a deep X-ray non-detection of the Superluminous SN 2015bn\footnote{Based on observations obtained with \textit{XMM-Newton}, an ESA science mission with instruments and contributions directly funded by ESA Member States and NASA.}}

\correspondingauthor{Kornpob Bhirombhakdi}
\email{kb291313@ohio.edu}

\author[0000-0003-0136-1281]{Kornpob Bhirombhakdi}
\affil{Astrophysical Institute, Department of Physics and Astronomy, 251B Clippinger Lab, Ohio University, 
Athens, OH 45701, USA}

\author[0000-0002-7706-5668]{Ryan Chornock}
\affil{Astrophysical Institute, Department of Physics and Astronomy, 251B Clippinger Lab, Ohio University, 
Athens, OH 45701, USA}

\author{Raffaella Margutti}
\affil{Center for Interdisciplinary Exploration and Research in Astrophysics (CIERA) and Department of Physics and Astronomy, Northwestern University, Evanston, IL 60208, USA}

\author{Matt Nicholl}
\affil{Harvard-Smithsonian Center for Astrophysics, 60 Garden Street, Cambridge, MA, 02138, USA}

\author{Brian D.~Metzger}
\affil{Department of Physics, Columbia University, New York, NY 10025, USA}

\author{Edo Berger}
\affil{Harvard-Smithsonian Center for Astrophysics, 60 Garden Street, Cambridge, MA, 02138, USA}

\author{Ben Margalit}
\affil{Department of Astronomy, University of California, Berkeley, CA 94720, USA}

\author{Dan Milisavljevic}
\affil{Department of Physics and Astronomy, Purdue University, 525 Northwestern Avenue, West Lafayette, IN 47907, USA}

\begin{abstract}

SN 2015bn is a nearby hydrogen-poor superluminous supernova (SLSN-I) that has been intensively observed in X-rays with the goal to detect the spin-down powered emission from a magnetar engine. The early-time UV/optical/infrared (UVOIR) data fit well to the magnetar model, but require leakage of energy at late times of $\lesssim 10^{43}$ erg s$^{-1}$, which is expected to be partially emitted in X-rays. Deep X-ray limits until $\sim$300 days after explosion revealed no X-ray emission. Here, we present the latest deep 0.3--10 keV X-ray limit at 805 days obtained with \textit{XMM-Newton}. We find $L_X < 10^{41}$ erg s$^{-1}$, with no direct evidence for central-engine powered emission. While the late-time optical data still follow the prediction of the magnetar model, the best-fit model to the bolometric light curve predicts that $\sim$97\% of the total input luminosity of the magnetar is escaping outside of the UVOIR bandpass at the time of observation. Our X-ray upper limit is $<$1.5\% of the input luminosity, strongly constraining the high-energy leakage, unless non-radiative losses are important. These deep X-ray observations identify a missing energy problem in SLSNe-I and we suggest future observations in hard X-rays and $\gamma$-rays for better constraints. Also, independent of the optical data, we constrain the parameter spaces of various X-ray emission scenarios, including ionization breakout by magnetar spin-down, shock interaction between the ejecta and external circumstellar medium, off-axis $\gamma$-ray burst afterglow, and black hole fallback accretion. 

\end{abstract}

\keywords{supernovae: individual (SN 2015bn) --- X-rays: individual (SN 2015bn)}

\section{Introduction} \label{sec:intro} 

\nocite{Nicholl2018.in.prep}

Superluminous supernovae (SLSNe) are known for being 10--100 times brighter at their UV/optical/infrared (UVOIR) peaks than typical SNe \citep{Chomiuk2011.slsn,Quimby2011.slsn,Gal-Yam2012}. They are either Type I for hydrogen poor or Type II for hydrogen rich. Power sources supplying their light curves, especially near peak light, are still uncertain. For SLSNe-I, the spinning-down magnetar scenario is currently the most favored explanation \citep{Kasen2010.magnetar,Woosley2010.magnetar,Chatzopoulos2013,Inserra2013.slsnI.magnetar,Nicholl2014.magnetar.slsnI,Metzger+15,Nicholl2017.mosfit,Wang2015,Wang2016}. However, there has been no direct evidence for the presence of a magnetar. One possible smoking gun would be the radio \citep{Murase2016.pwn,Omand2018.radio.slsn,Margalit&Metzger18}, or X-ray \citep{Kotera2013.PWN.late.time,Metzger2014,Metzger2014.Piro} emission resulting from the cooling of the relativistic particles accelerated in the pulsar wind nebula (PWN) that is created by the magnetar. 

The timescale for X-ray emission from the engine to emerge from the ejecta in SLSNe-I is still theoretically uncertain. The X-ray photons either escape without strongly affecting the ionization state of the ejecta (i.e., leakage; \citealt{Wang2015,Margalit2018.cloudy}), or by heating and ionizing the ejecta until they become transparent to the emission (i.e., ionization breakout; \citealt{Metzger2014,Metzger2014.Piro,Margalit2018.cloudy}). In the leakage scenario, the expansion of the ejecta causes it to become transparent to X-ray photons (dilution effects; \citealt{Margalit2018.cloudy}) with a timescale of $\gtrsim$100 years, while the ionization breakout has a much shorter timescale \citep{Metzger2014}. The X-ray searches during the past decade, covering up to $\sim$5 years post explosion from various events (see \citealt{Margutti2017.X-ray.SLSNI} for the compilation) have led to non-detections, except for SCP06F6 \citep{Levan2013.scp06f6}.

SN 2015bn is one of the closest, and best studies SLSNe-I \citep{Nicholl2016.2015bn.early,Nicholl2016slsn2015bnNebular,Jerkstrand2017.2015bn.spectral.syn.oxygen}, providing the opportunity to perform deep X-ray searches at ages $>$1 year \citep{Nicholl2016slsn2015bnNebular,Inserra2017lsq14an,Margutti2017.X-ray.SLSNI}. The latest previous X-ray observation was a non-detection with $L_X$(0.3--10 keV) $\lesssim 10^{41}$ erg s$^{-1}$ at $\sim$300 days after explosion \citep{Margutti2017.X-ray.SLSNI}. Here, we present an additional X-ray observation at 805 days, and discuss the implications for the power source of the SN.  A complementary paper by Nicholl et al. (in prep.) discusses the constraints provided by late-time optical observations. Throughout, we apply the redshift $z = 0.1136$, the luminosity distance 513 Mpc, and the explosion date MJD 57013 \citep{Margutti2017.X-ray.SLSNI}. Any calendar date refers to Universal Time. Also, the supernova phases, or ages, are measured since the explosion in the rest frame, unless specified otherwise.

\section{Data} \label{sec:data} 

\begin{figure}
  \centering
  \includegraphics[width=0.45\textwidth, angle=0]{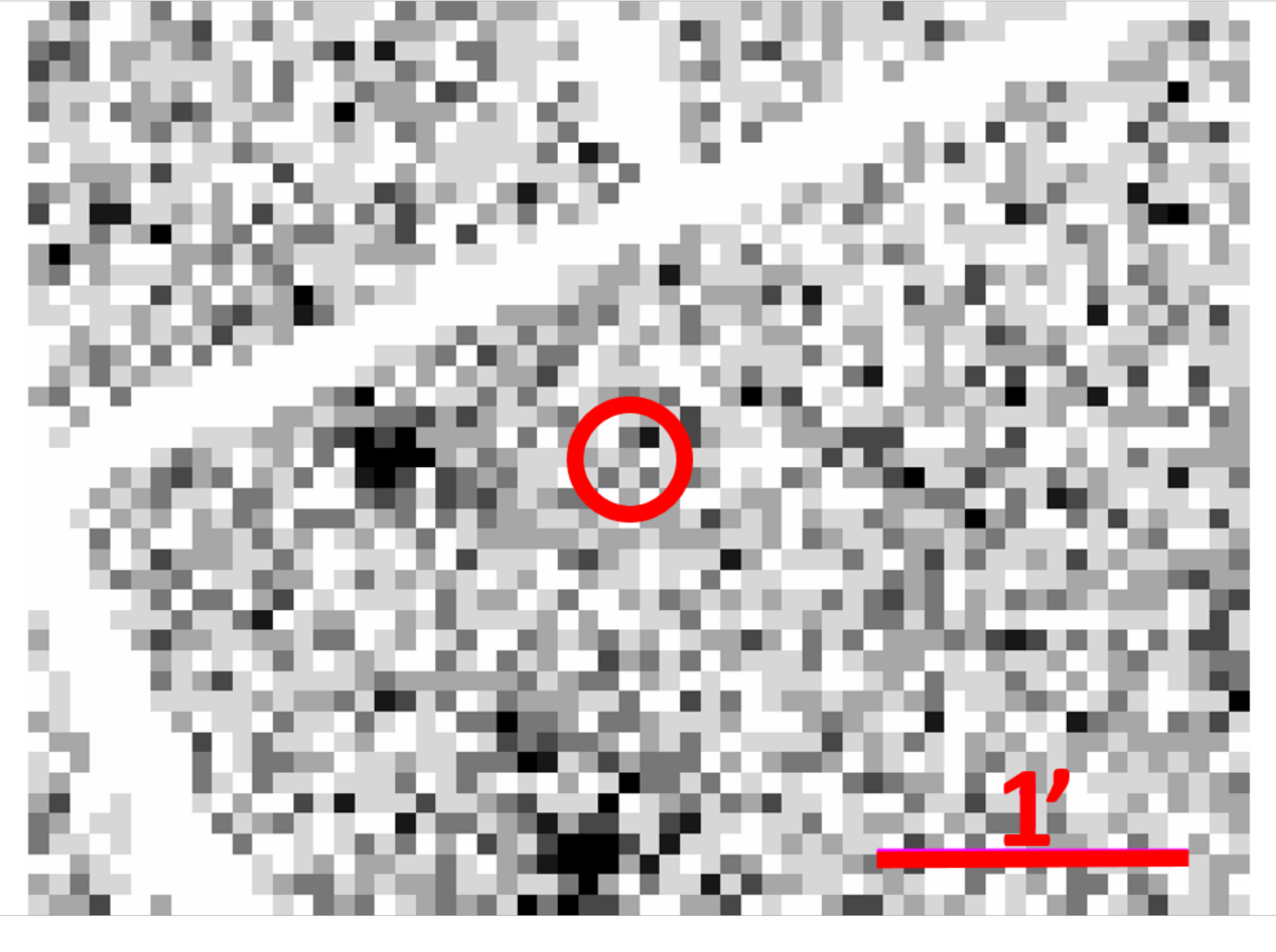}
  \caption{EPIC-pn image of SLSN 2015bn (10\arcsec\ red circle) in 0.3--10 keV X-rays at 805 days. Black = high counts. North is up and east is to the left. The red scale bar is 1\arcmin\ in length.}
  \label{fig:2015bn}
\end{figure}

SN 2015bn is located at $\alpha = $ 11h 33m 41.551s, $\delta =$ +00d 43m 33.40s (J2000.0) \citep{Nicholl2016.2015bn.early,Nicholl2016slsn2015bnNebular,Margutti2017.X-ray.SLSNI}. Our latest observation includes one epoch of X-ray photometry from the European Photon Imaging Camera of the European Space Agency's X-ray Multi-Mirror Mission (\href{https://www.cosmos.esa.int/web/xmm-newton}{\textit{XMM-Newton}}\footnote{https://www.cosmos.esa.int/web/xmm-newton}; ID: 0802860201; PI: Chornock),  in both Metal Oxide Semi-conductor (MOS) 1, MOS2, and pn cameras \citep{Struder2001.EPIC.pn,Turner2001.EPIC.mos}. The observation started on 2017 June 5 (MJD 57909) and ended on 2017 June 6 (MJD 57910), at phase $\sim$805 days. The most constraining image is from EPIC-pn with the thin filter and 37.7 ks of exposure, so all subsequent analysis is performed on this image. By applying the Science Analysis System (\href{https://www.cosmos.esa.int/web/xmm-newton/sas-news}{SAS}\footnote{https://www.cosmos.esa.int/web/xmm-newton/sas-news}, version 20170719\_1539-16.1.0), and following the standard procedure for image reduction, the data have a Good Time Interval (GTI) of 35.7 ks. 

As shown in \autoref{fig:2015bn}, no X-ray source is detected at the location of the SN. The 3$\sigma$ upper limit is estimated to be $1.57 \times 10^{-3}$ count s$^{-1}$ in the 0.3--10 keV bandpass. By applying \webpimms \footnote{https://heasarc.gsfc.nasa.gov/Tools/multimissiontools.html}, and using a Galactic neutral hydrogen column density in the direction of the transient of $NH_{\rm{MW}}=2.4 \times 10^{20}$ cm$^{-2}$ \citep{Kalberla2005.hydrogen.map}, and assuming zero intrinsic column density of neutral hydrogen, the upper limit on the unabsorbed flux (0.3--10~keV) is $3.6 \times 10^{-15}$ erg s$^{-1}$ cm$^{-2}$ ($L_X \lesssim 1.1 \times 10^{41}$ erg s$^{-1}$) assuming a power law spectrum with photon index $\Gamma = 2$, or $5.3 \times 10^{-15}$ erg s$^{-1}$ cm$^{-2}$ ($L_X \lesssim 1.7 \times 10^{41}$ erg s$^{-1}$) assuming a 20 keV thermal bremsstrahlung model (this flux conversion is insensitive to the precise temperature as long as it is above the \textit{XMM} bandpass).  

\section{Analysis and Discussion} \label{sec:ana} 

The X-ray non-detections of SLSN can provide constraints on the explosion's properties and the properties of its environment (see \citealt{Margutti2017.X-ray.SLSNI} for examples). 
In this section, four X-ray emitting scenarios are considered: magnetar spin-down (section \ref{ssec:mag}), ejecta-medium interaction (section \ref{ssec:ambient}), off-axis GRB afterglows (section \ref{ssec:grb}), and black hole (BH) fallback accretion (section \ref{ssec:bh}).

\subsection{Constraining magnetar spin-down} \label{ssec:mag} 

Magnetar spin-down \citep{Kasen2010.magnetar,Woosley2010.magnetar} is the most favored explanation for SLSNe-I currently. The magnetar, which is a neutron star with the surface dipole magnetic strength ${>} 10^{13}$ G, releases its rotational energy from magnetic braking \citep{Duncan1992}, creating a PWN which is composed of energetic electron/positron pairs \citep{Gaensler2006.pwn}. The particles cool down by synchrotron or inverse Compton emission, which in turn creates more pairs if the energy allows, resulting in the pair cascade \citep{Lightman1987.pair.cascade,Svensson1987.pair.cascade,Vurm2009.pair.cascade}. X-ray photons are emitted but may not emerge from the ejecta due to photoelectric absorption.
A recent example of this may be SN 2012au, whose 6-year optical spectrum showed evidence for ionization of oxygen by a PWN, but X-ray observations resulted in a non-detection, which was interpreted as being due to high ejecta opacity \citep{Milisavljevic2018.2012au.spectrum.6yr}.
Reprocessing of this absorbed emission by the ejecta is responsible for powering the optical/UV light (\citealt{Metzger2014}).

In this section, we compare the observed energy to the predicted input (section \ref{sssec:lc}), then constrain the parameter space of magnetar spin-down under the X-ray ionization breakout scenario (section \ref{sssec:breakout}; \citealt{Metzger2014}), and last discuss the possibility of observing the breakout in the future (section \ref{sssec:future}).

\subsubsection{Light curve in magnetar spin-down scenario} \label{sssec:lc} 

\begin{figure}
  \centering
  \hspace*{-1.4in}
  \includegraphics[width=0.45\textwidth, angle=90]{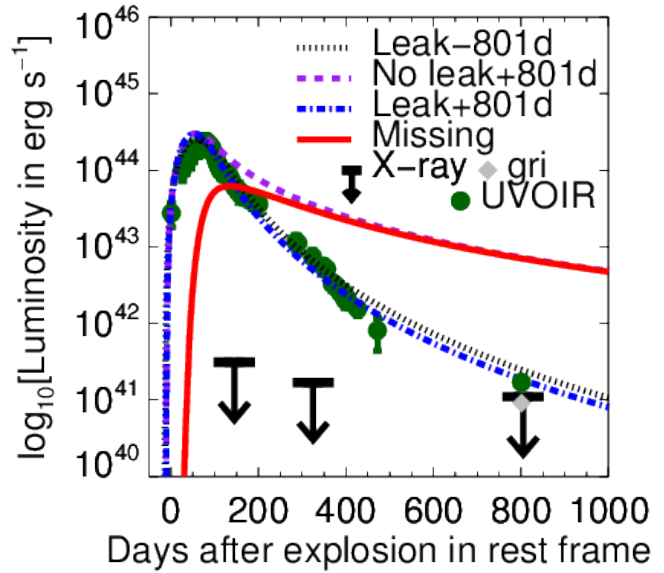}
  \caption{Light curve of SN 2015bn. Dark green dots = UVOIR data ($<$801 days from \citealt{Nicholl2016.2015bn.early,Nicholl2016slsn2015bnNebular} and at 801 days from Nicholl et al., in prep.). Black arrows = 3-sigma upper limits from 0.3--10 keV X-ray observations from \textit{XMM-Newton} \citep{Margutti2017.X-ray.SLSNI}. Gray diamond = $gri$ luminosity at 801 days (Nicholl et al., in prep.). Black dotted line = magnetar spin-down model with leakage effects without including the 801-day data \citep{Nicholl2017.mosfit}. Purple dashed line = magnetar spin-down model without leakage effects and including the 801-day data (Nicholl et al., in prep.). Blue dot-dashed line = magnetar spin-down model with leakage effects and including the 801-day data (Nicholl et al., in prep.). Red solid line = the difference in luminosity between the models with and without leakage, representing the missing energy. These observations identify a missing energy problem in SLSNe-I.
}
  \label{fig:discrepancy}
\end{figure}

\begin{table*}
	\centering
    \caption{Expected luminosity in various scenarios}
    \label{table-luminosity}
    \begin{tabular}{lcccrrrrrr}
		\hline
        \multicolumn{1}{c}{Case} & \multicolumn{1}{c}{Model (M) or} & \multicolumn{1}{c}{Include} & \multicolumn{1}{c}{Bandpass} & \multicolumn{6}{c}{Luminosity ($10^{42}$ erg s$^{-1}$)} \\
         & \multicolumn{1}{c}{observation (O)?} & \multicolumn{1}{c}{leakage effects?} & & \multicolumn{2}{c}{145 days} & \multicolumn{2}{c}{325 days} & \multicolumn{2}{c}{805 days} \\
         \hline
         No leak+801d & M & N & Total & 125.17 & & 35.34 & & 7.22 & \\
         Leak+801d & M & Y & Total & 63.52 & 51\% & 5.03 & 14\% & 0.18 & 2.5\% \\
         UVOIR data & O & - & UVOIR & 60.67 & 48\% & 7.44 & 21\% & 0.17 & 2.4\% \\
         X-ray data & O & - & 0.3--10 keV & ${<} 0.31$ & $<$0.2\% & ${<} 0.17$ & $<$0.5\% & ${<} 0.11$ & $<$1.5\% \\
         \hline
	\end{tabular}
\end{table*}

\autoref{fig:discrepancy} shows the light curve of SN 2015bn, along with fits to the magnetar model. The latest optical $gri$ luminosity observed on 2017 June 1 (MJD 57905), corresponding to phase 801 days, is from Nicholl et al. (in prep.), with $L_{UVOIR} \approx 1.7 \times 10^{41}$ erg~s$^{-1}$. The fit lines are the total bolometric luminosity from the ``slsn'' model, which is the modified magnetar spin-down model \citep{Nicholl2017.mosfit} of \mosfit \footnote{https://mosfit.readthedocs.io/en/latest/} \citep{Guillochon2017.mosfit}. We note that the ``Leak-801d'' model was presented by \cite{Nicholl2017.mosfit}, and was downloaded from The Open Supernova Catalog (\OSC; \citealt{guillochon.osc})\footnote{https://sne.space/}. Moreover, the ``No leak+801d'' model is estimated by applying the same parameters from the fit of the ``Leak+801d'' but changing the leakage coefficient (see \citealt{Chatzopoulos2013,Wang2015,Nicholl2017.mosfit} about the leakage effect) so that the leakage effect is negligible. The ``Missing'' line shows the difference between the models with and without leakage.

As presented in the figure, adding the 801-day UVOIR data into the fit does not significantly change the fit parameters: initial spin-down period 2.16 ms, magnetic field strength $3 \times 10^{13}$ G, and ejecta mass 11.7 \mdot\ for the median values \citep{Nicholl2017.mosfit}. We note that there are other magnetar spin-down results in literature \citep{Nicholl2016.2015bn.early,Nicholl2016slsn2015bnNebular} which have similar parameters, but we include only the ones from \mosfit\ for consistency. Moreover, the modified magnetar spin-down in \mosfit, ``slsn'', has more parameters than mentioned here (see \citealt{Nicholl2017.mosfit}), but those extra parameters are irrelevant to the discussion. 

For the case without the leakage effect, which represents the efficient conversion of the total spin-down luminosity into radiation \citep{Chatzopoulos2013,Wang2015}, the discrepancy with the UVOIR observations has started since about 100 days, corresponding to its spectrum starting to show some noticeable changes \citep{Nicholl2016.2015bn.early}. Then, the gap tends to increase with age while the SN evolved into the nebular phase. \autoref{table-luminosity} numerically shows the discrepancy at the three epochs corresponding to the deep \textit{XMM-Newton} observations. The percentage of the luminosity relative to that of the non-leakage case is also calculated. 

The models imply that the leakage continuously increases relative to the total luminosity (i.e., $\sim$50\% at 145 days to $\sim$97\% at 805 days). The X-ray non-detections mean that radiation in the 0.3--10 keV bandpass cannot account for the total leakage.
We have three possibilities: non-radiative losses (e.g., adiabatic expansion and accelerating ejecta due to the expanding hot bubble from the PWN's activity, or simply losing non-interacting particles created from the PWN's activity), radiative losses outside our observational bands, or that the magnetar model is not correct.

Since the magnetar injects relativistic particles and high-energy photons (i.e., X-ray and $\gamma$-ray) into the PWN, the energy has to be converted to the UVOIR and soft X-ray photons that we observe. If this energy can escape the ejecta at other wavelengths (such as the $\gamma$-rays), the observed bandpasses might not provide a complete account of the bolometric luminosity.
Also, the \mosfit\ model assumes a blackbody SED in the optical/infrared bandpass, which might not be accurate during the nebular phase due to strong line emission. Furthermore, it is also possible that the magnetar fit to the peak of the light curve might not apply at late times if, for instance, the spin-down parameters change due to accretion \citep{Metzger2018.fallback}, or if the magnetar collapses to a black hole \citep{Moriya2016.SLSN-I.magnetar.to.blackhole}.

Last, we also note that the analysis is sensitive to assumptions implicitly included in the leakage term (e.g., homologous expansion and constant leakage coefficient). The assumption of spherical symmetry is vital and might not be accurate in some scenarios such as having clumps or jets. Moreover, the analysis assumes no emission is contributed via other mechanisms such as radioactivity, circumstellar interaction, or a light echo (such as that observed in the SLSN-I iPTF16eh; \citealt{Lunnan2018.slsnI.echo}). However, the late time optical observations of SN 2015bn strongly constrain these mechanisms (Nicholl et al., in prep.).

\subsubsection{X-ray ionization breakout} \label{sssec:breakout} 

\begin{figure}
  \centering
  \hspace*{-0.3in}
  \includegraphics[width=0.45\textwidth]{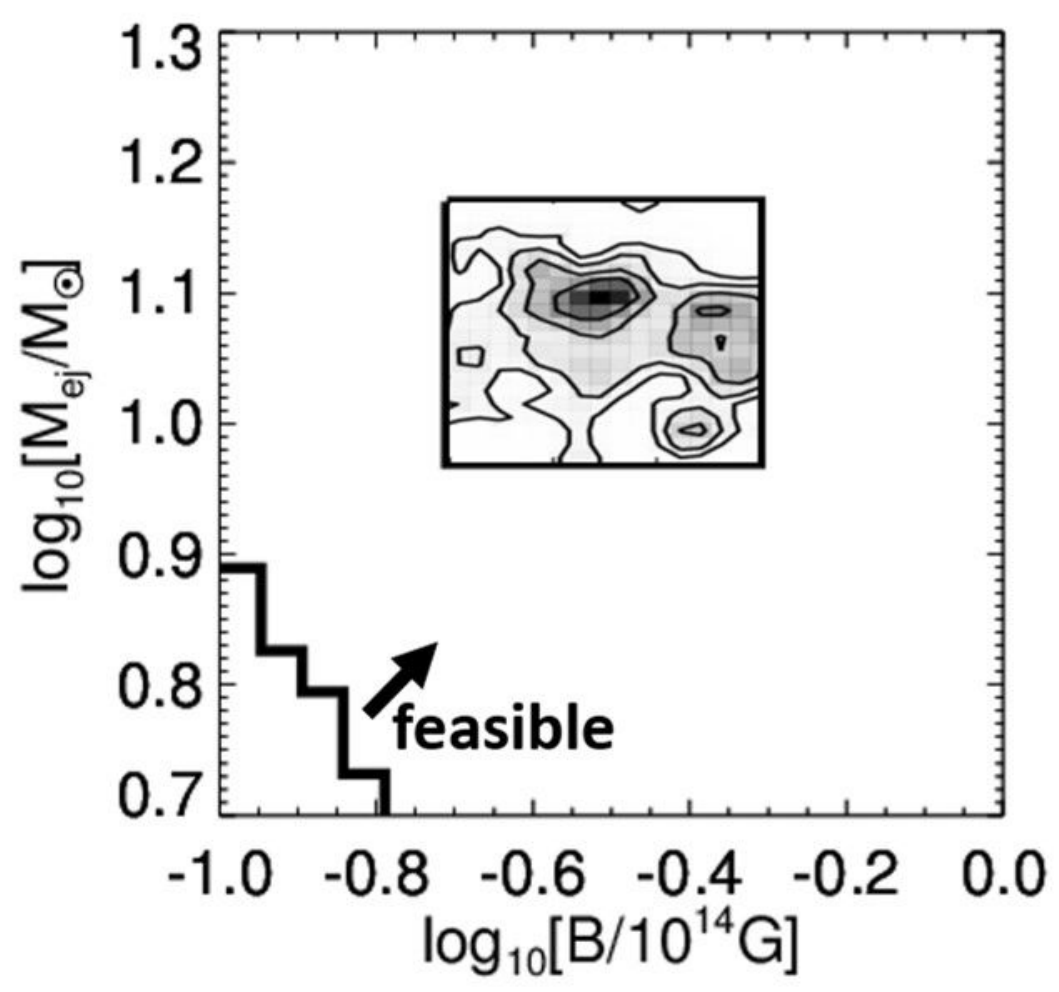}
  \caption{Allowed parameter space, assuming that X-ray ionization breakout will occur after 805~d and that $T_e = 10^5$ K. The area to the right of the line is feasible. The rectangular area with the contours approximately corresponds to the posterior distribution estimated from the UVOIR data by \citet{Nicholl2017.mosfit} and is entirely feasible.
  }
  \label{fig:1e5K.v15000}
\end{figure}

We constrain the parameter space of the magnetar spin-down in this section by applying the model of the X-ray ionization breakout \citep{Metzger2014}. Because of the frequent X-ray observations during early times (see \citealt{Margutti2017.X-ray.SLSNI} for the compilation), we are safe to conclude that the X-ray ionization breakout has not happened in the past; if breakout had happened and the X-ray light curve had behaved as specified by the model (i.e., $L_X \propto t^{-2}$), we should have detected it with X-ray luminosity $L_X > 10^{41}$ erg s$^{-1}$ at some epochs before 805 days. These non-detections through 805 days are consistent with the predictions that the breakout timescales are $\sim$1--100 years \citep{Inserra2017lsq14an,Margutti2017.X-ray.SLSNI}. In the following analysis, we assume that X-ray ionization breakout will take place in the future, and that the model remains valid to these late times. (We discuss possible caveats in the following section, \ref{sssec:future}.)

The timescale for the X-ray ionization breakout is estimated by following the model in \cite{Metzger2014} (see also \citealt{Margutti2017.X-ray.SLSNI}, equations 2, 4, and 5). Since SN 2015bn is oxygen dominated, we are interested in the breakout of oxygen ($Z = 8$). We assume the mass fraction of oxygen in the ejecta $X_O = 0.7$ \citep{Nicholl2016slsn2015bnNebular,Jerkstrand2017.2015bn.spectral.syn.oxygen}, the characteristic ejecta velocity $v_{ej} = 10^4$ km s$^{-1}$, and the electron temperature $T_e = 10^5$ K corresponding to the temperature for ionizing oxygen \citep{Metzger2014,Margutti2017.X-ray.SLSNI}. We constructed a grid of ejecta mass ($M_{ej}$) and magnetic field strength ($B$) in the ranges 5--20 $\textrm{M}_\odot$ and\ $10^{13}$--$10^{14}$~G.  The inferred breakout timescales range over $\sim$1--$10^3$ years. Then, we identify ``feasible'' grids if the timescale is $>$805 days. \autoref{fig:1e5K.v15000} shows the result with the contour of the posterior distribution (in the rectangular area) presented by \cite{Nicholl2017.mosfit}, which is estimated from the UVOIR data. The X-ray non-detections, independently of the UVOIR data, rule out the parameter space of the magnetar spin-down with low ejecta mass ($\lesssim$8 \mdot) and low magnetic strength (${\lesssim} 2 \times 10^{13}$ G). The feasible space is consistent with, but less constrained than, that of the UVOIR data. 

In summary, we demonstrate how even non-detections in the X-rays can constrain magnetar spin-down independently of the UVOIR data. For SN 2015bn, the X-ray non-detections until 805 days can rule out a portion of the parameter space of the magnetar spin-down with low ejecta mass and low magnetic strength. Later epochs of X-ray observation, if still non-detections, will shift the feasible line to the right, possibly ruling out some overlapping space with the results from the fits to the UVOIR data.

We note that the electron temperature is uncertain and can significantly affect the analysis. Although the characteristic temperature in the PWN is ${\sim} 10^7$ K \citep{Metzger2014}, the temperature of gas in the ionized layers of the ejecta, $T_{e}$, is significantly less than this. Here we have assumed a gas temperature $T_{e} = 10^5$ K \citep{Metzger2014}, but the actual temperature could be lower than this at very late times \citep{Margalit2018.cloudy}. Since the ionization breakout timescale obeys $t_{ion} \propto T_e^{-n}$ for $n = \{0.3,0.8\}$ depending on some conditions \citep{Metzger2014}, this can increase the break-out time, implying a large shift of the allowed parameter space comapared to that shown in \autoref{fig:1e5K.v15000}. Indeed, \citet{Margalit2018.cloudy} find that X-ray ionization breakout is unlikely to occur at late times in SLSNe, due in part to the decreasing ejecta temperature (increasing recombination rate) as the ejecta expands.

\subsubsection{X-ray ionization breakout in the future?} \label{sssec:future} 

In the magnetar-powered PWN, energetic electron/positron pairs cool, creating gamma-ray photons, which can further annihilate and create lower energetic pairs, which then can Compton upscatter the nebular radiation \citep{Metzger2014}.  This process, which repeats multiple times, is known as a ``pair cascade'' \citep{Svensson1987.pair.cascade}. If the system is sufficiently ``compact" (sufficiently high energy density), the process becomes ``saturated" after many cycles, resulting in flat photon spectral energy distribution (SED), with $F_{\nu} \propto \nu^{-\beta}$ and $\beta \sim 1$. Otherwise, the SED from synchrotron or Inverse Compton emission is likely to be harder, $\beta \lesssim 1$, and therefore the X-ray emission will be fainter than predicted by the model \citep{Metzger2014}.

The ionization breakout process requires a large density of UV/X-ray photons and thus favors a relatively soft nebula spectra (high compactness).  
In the magnetar scenario, as the ejecta expands, the nebula compactness drops.  For SN 2015bn, we estimate the compactness at 805 days to be ${\sim} 10^{-3}$ (see equation 13 in \citealt{Metzger2014} and 4 in \citealt{Margutti2017.X-ray.SLSNI} with parameters in \citealt{Nicholl2017.mosfit}), given the albedo 0.5 and the diffusive timescale ${\sim}$80 days (approximately the rising time of the UVOIR light curve; \citealt{Arnett1980,Arnett1982,Chatzopoulos2012}).  Such low compactness means that in principle high energy gamma-rays could escape from the nebula (without creating pairs) and thus leaving few UV/X-ray photons to ionize the ejecta.  Future studies of ionization break-out, analogous to those of \citet{Margalit2018.cloudy} should account self-consistently for the predicted hardening in the ionizing spectrum at late times.

\subsection{Constraining ejecta-medium interaction} \label{ssec:ambient} 

\begin{figure}
  \centering
  \hspace*{-1.4in}  
  \includegraphics[width=0.45\textwidth, angle=90]{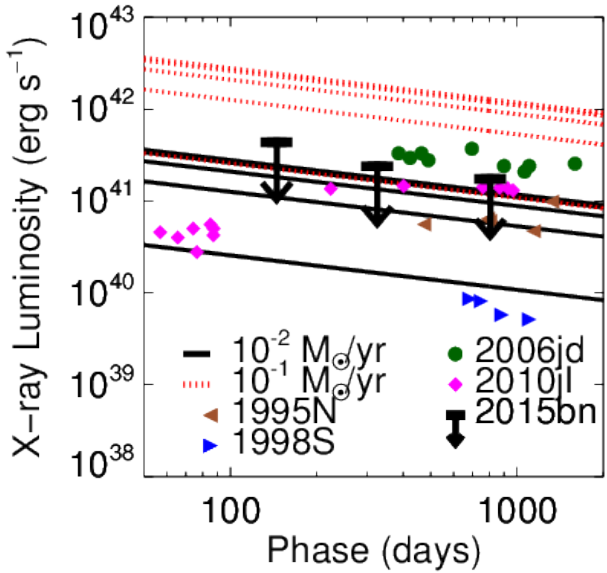}
  \caption{X-ray luminosity (0.3--10 keV) with predicted lines from the ejecta-medium interaction models. Black arrow = 3$\sigma$ upper limits of X-ray data of SN 2015bn from \textit{XMM-Newton}, assuming zero intrinsic absorption and 20 keV thermal bremsstrahlung model. Lines = predicted luminosity from the reverse shock in the interaction model \citep{Fransson1984.csi}, assuming $v_w = 10^3$ km s$^{-1}$, and $\dot{M} = 10^{-1}$ (red dotted), $10^{-2}$ (black solid) \mdot\ yr$^{-1}$ with the intrinsic column density of neutral hydrogen of $10^{20},10^{21},10^{22},10^{23},10^{24}$ cm$^{-2}$ (from top to bottom). X-ray data for some SNe IIn are presented, including SN 1995N (brown leftwards triangle; \citealt{Chandra2005.1995N.X-ray}), SN 1998S (blue rightwards triangle; \citealt{Pooley2002.1998S.1999em.X-ray}), SN 2006jd (dark green circle; \citealt{Chandra2012.2006jd.X-ray.radio}), and SN 2010jl (magenta diamond; \citealt{Ofek2014.2010jl}).}
\label{fig:ambient}
\end{figure}

X-ray emission in the ejecta-medium interaction is well studied in many events, especially SNe IIn like SN 1998S \citep{Pooley2002.1998S.1999em.X-ray}, SN 2006jd \citep{Chandra2012.2006jd.X-ray.radio}, and SN 2010jl \citep{Chandra2015.2010jl.xray}, and SNe Ib/c \citep{Chevalier2006.CSI.Ib.Ic}. In this scenario, the X-ray emission constrains the medium density at the location of the shock, in our case $\sim  10^{17}$~cm from the explosion site at 805 days after explosion, given $10^4$ km s$^{-1}$ for the typical shock velocity (see \citealt{Margutti2017.X-ray.SLSNI} for the constraints on the medium density at earlier epochs). Even though there has been no clear sign of circumstellar interaction during the earlier phases \citep{Nicholl2016.2015bn.early,Nicholl2016slsn2015bnNebular}, the medium at the late phases might have different properties. There has been growing evidence for hydrogen-poor SNe showing hydrogen features from the interaction in their late-time spectra \citep{Milisavljevic2015.2014C.Ic.to.IIn,Yan2015,Yan2017.slsnI.lateH,Chen2018.2017ens.SLSN-I.to.IIn,Kuncarayakti2018.2017dio.Ic.to.IIn,Mauerhan2018.2004dk.Ic.to.IIn}, and there is the recent evidence of the light echo from iPTF16eh \citep{Lunnan2018.slsnI.echo} implying a significant amount of hydrogen-poor circumstellar medium in a SLSN-I at ${\sim} 10^{17}$ cm. Moreover, the early-time undulations seen in the optical light curve of SN 2015bn \citep{Nicholl2016.2015bn.early,Nicholl2016slsn2015bnNebular} might imply inhomogeneities in the circumstellar medium. Therefore, estimating the medium properties at various phases can help constrain the presence of interaction. 

In the absence of more detailed simulations, we do not know what the main emission mechanism for the X-ray photons from the ejecta-medium interaction would be at this epoch. At earlier epochs, inverse Compton scattering dominates the emission \citep{Margutti2017.X-ray.SLSNI}. At late times, synchrotron radiation dominates the non-thermal X-ray emission unless the medium is sufficiently dense, in which case the emission is thermal bremsstrahlung \citep{Chevalier2017.inbook.thermal.nonthermal}. The estimates here assume the latter scenario, and also assume that the soft 0.3--10 keV X-ray emission is dominated by the reverse shock according to its characteristic temperature \citep{Fransson1984.csi,Chevalier2017.inbook.thermal.nonthermal}, as expected in a medium with the density profile of a wind. Since we also cannot tell whether the X-ray photons can escape the dense reverse shock from its absorption, our estimation here presents a conservative upper limit.

We apply the model from \cite{Fransson1984.csi} (see also equation 16 in \citealt{Chevalier2017.inbook.thermal.nonthermal}). Since, under this assumption, the emission is likely to be thermal, in this section we estimate the emission by assuming a 20 keV thermal bremsstrahlung model, representative of detections of previous strongly interacting SNe (e.g., \citealt{Chandra2015.2010jl.xray,Margutti2017.2014C.X-ray}).
All of the calibration was estimated using \webpimms. \autoref{fig:ambient} presents the absorbed luminosity of the three \textit{XMM-Newton} X-ray data points, only correcting for the Galactic ($NH_{\mathrm{gal}}$) column density of neutral hydrogen of $2.37 \times 10^{20}$ cm$^{-2}$. We note that the assumed zero intrinsic absorption ($NH_{\mathrm{int}}$) gives us the conventional lower limit of the luminosity, since more intrinsic absorption shifts the limit to higher luminosity (given a fixed count rate). We also assume a steady wind environment.

The absorbed luminosity ($L$) depends on the mass loss rate ($\dot{M}$), steady wind velocity ($v_w$), the ejecta velocity ($v_{ej}$), the power-law index of the density of the outer part of the ejecta ($n$), the absorption parameters (i.e., $NH_{\mathrm{gal}}$, and $NH_{\mathrm{int}}$), and the reference day for scaling (i.e., $L \propto t^{-3 / (n-2) }$). We use $n = 10$ as the typical value for a stripped-envelope SN \citep{Chevalier2017.inbook.thermal.nonthermal}, $v_w = 10^3$ km s$^{-1}$, $v_{ej} = 10^4$ km s$^{-1}$, $NH_{\mathrm{gal}} = 2.37 \times 10^{20}$ cm$^{-2}$, and the scaling relative to 805 days. \autoref{fig:ambient} shows the models with $\dot{M} = 10^{-2}$, and $10^{-1}$ \mdot\ yr$^{-1}$. Each model is estimated with $NH_{\mathrm{int}} = 10^{20},10^{21},10^{22},10^{23},10^{24}$ cm$^{-2}$ (from top to bottom). Since $L \propto \dot{M} v_{ej}^3$, the X-ray data are consistent with the model $\dot{M} < 10^{-2}$ \mdot\ yr$^{-1}$, $v_{ej} < 10^4$ km s$^{-1}$, and any $NH_{\mathrm{int}}$. For a larger mass loss rate, the data might be consistent with the predictions if the intrinsic absorption is large. This result is also consistent with the radio limits at late times (Nicholl et al., in prep.).

\autoref{fig:ambient} also shows some SNe IIn (see \citealt{Dwarkadas2012.X-ray.IIn} and references therein) with soft X-ray (0.3--10 keV) detections at comparable ages to SN 2015bn. The data demonstrate that the X-ray luminosity in some strongly interacting SNe IIn (e.g., SN 2006jd; \citealt{Chandra2012.2006jd.X-ray.radio}) can be brighter than the upper limits for SN 2015bn.

\subsection{Off-axis GRB} \label{ssec:grb} 

Some SNe with features similar to SLSNe might also harbor jets, like the luminous SN 2011kl associated with GRB 111209A \citep{Greiner2015.2011kl.ulgrb,Margalit2018.grb.slsn.frb}. X-ray to radio emission can be observed at late times after the explosion from the jet interaction with the circumburst medium \citep{Nousek2006.X-ray.afterglow,Roming2009.UV.optical.afterglow,Chandra2012.radio.afterglow}. Depending mainly on the injected energy, the medium properties, the energy conversion factors, the jet opening angle, and the angle between the line of sight and the jet axis, the afterglows vary in the timescale and the SED \citep{Granot2002.afterglow,Granot2002.afterglow.spectrum}. 

For SN 2015bn, the earlier X-ray and radio non-detections rule out portions of the parameter space \citep{Nicholl2016.2015bn.early,Margutti2017.X-ray.SLSNI,Coppejans2018.radio.grb.slsnI}. Here, we apply the same BOXFIT \citep{vanEerten2012.afterglow.simulations} simulated 0.3--10 keV X-ray light curves in the scenario of off-axis GRB jets, as presented by \cite{Margutti2017.X-ray.SLSNI}, with our latest X-ray non-detection. The data further rule out only a very small additional portion of parameter space, including most cases of jets with unrealistically high isotropic equivalent kinetic energy ${>} 10^{55}$~erg, the circumburst medium with ${>} 10^{-3}$ cm$^{-3}$ uniform density profile, the jet opening angle $<$15$^{\circ}$, and the line of sight angle $<$30$^{\circ}$ with respect to the jet axis, given the fiducial values of the energy conversion factors: $\epsilon_B = 0.01$ and $\epsilon_e = 0.1$.

\subsection{Black hole as a central engine} \label{ssec:bh} 

Instead of forming a neutron star, a SLSN-I might form a BH, in which case the UVOIR peak would be powered by the fallback accretion of slow ejecta at the inner boundary \citep{Dexter2013}. Although this is unlikely to be the case for SN 2015bn due to the large accreted mass required to power the main UVOIR peak \citep{Moriya2018.fallback.slsnI.MOSFiT}, a BH could also form at late times from a magnetar accreting enough fallback material \citep{Moriya2016.SLSN-I.magnetar.to.blackhole}. In either case, X-rays could be emitted as the result of the central engine's activity.
Our late-time X-ray limit constrains such a scenario. The combined UVOIR and X-ray data at $\sim$800 days imply that the bolometric luminosity is ${\lesssim} 100$ times  the Eddington value for a central BH with mass 10~M$_{\odot}$, although the fraction of the accretion luminosity to escape would depend on the ionization state and amount of soft X-ray absorption in the ejecta, as discussed above in the magnetar scenario.

\section{Conclusion} 

We present the latest deep X-ray observation from \textit{XMM-Newton} of SN 2015bn, one of the closest SLSNe-I. The observation corresponding to the phase 805 days shows a 0.3--10 keV X-ray non-detection, with a 3-sigma upper limit of $L_X \lesssim 10^{41}$ erg s$^{-1}$, with the implication that we still cannot distinguish models for the power source of the event. In the magnetar spin-down scenario, the best-fit model predicts $\sim$97\% of the total energy input leaks outside of the UV/optical/infrared bandpass, and the UVOIR data up to $\sim$800 days follow the prediction. Our X-ray upper limit is $<$1.5\% of the total, strongly constraining the leakage, unless non-radiative loss is important. 

Independent of the UVOIR data, the X-ray upper limits rule out the possibility of having an ionization breakout earlier than 805 days, and rule out magnetar spin-down with low ejecta mass ($\lesssim$8 \mdot) and low magnetic strength (${\lesssim} 2 \times 10^{13}$ G), consistent with the results from the UVOIR data in recent literature \citep{Jerkstrand2017.2015bn.spectral.syn.oxygen,Nicholl2017.mosfit}. In the future, however, the breakout is unlikely to happen due to the compactness problem. This issue is generally true for any old-age SNe. In the ejecta-medium interaction scenario, we constrain the environment at ${\sim} 10^{17}$ cm to be ${\lesssim} 10^{-2}$ \mdot\ yr$^{-1}$ for a $10^3$ km s$^{-1}$ steady wind. In the off-axis GRB and BH fallback scenarios, our observations only rule out extreme models.

We note that the analysis here is sensitive to some assumptions. For example, the SED estimated in the ionization breakout model, which assumes the pair-cascade saturation that seems true at young ages, might not be valid in the low-compactness regime at old ages. 
In this regime, we note that the SEDs are expected to be harder than assumed in the ionization breakout model \citep{Metzger2014}, and therefore X-ray emission should be fainter than predicted and observing the emission will be challenging. 
The feasible line presented in \autoref{fig:1e5K.v15000} is also sensitive to the electron temperature at the ionizing layers. The magnetar spin-down model, which assumes some parameters to be constants since early times and includes the leakage effects with a constant coefficient, might not be accurate at old ages. The estimated density of the ambient medium in the interaction scenario assumes the X-ray emission is dominated by the reverse shock. All models assume spherical symmetry, which might not hold \citep{Inserra2016.2015bn.polarimetry,Leloudas2017.polar.2015bn}.

The search for the smoking gun of a central engine is still ongoing. Nicholl et al. (in prep.) suggests that the late-time flattening of the optical light curve of SN~2015bn after $\sim$500 days with a decline rate slower than that of $^{56}$Co decay is evidence for the continuous input of energy from a central engine, although confirmation requires more examples. In addition, the energetic SN Ib-pec 2012au, which might be a lower-luminosity counterpart of some SLSNe-I \citep{Milisavljevic2013.2012au.early}, including SN 2015bn \citep{Nicholl2016slsn2015bnNebular},
had an optical spectrum at an age of 6~yr that was recently interpreted as photoionized oxygen-rich gas shocked by a high pressure PWN \citep{Milisavljevic2018.2012au.spectrum.6yr}.
For the X-ray signal, we still encourage the early-time observations, despite many non-detections in the past, because there is a chance of observing the signal similar to what was observed in SCP06F6 \citep{Levan2013.scp06f6}.  Asphericity might play a significant role in the observed signal, which yields an additional opportunity to study the geometric distribution of the explosion. 

According to \cite{Margalit2018.cloudy}, the early-time ionization breakout timescale is less than the spin-down timescale, typically $<$1 year. Therefore, this might be the golden period to observe such the scenario. After the first year, the chance of observing ionization breakout is low, but still possible. We also suggest observations in MeV--GeV $\gamma$-rays to constrain the high energy emission, as might be the case for direct leakage from the PWN in the low-compactness regime. We note that recent {\it Fermi}-Large Area Telescope observations of SN 2015bn set a limit on the $>$600 MeV $\gamma$-ray luminosity of $L_\gamma \lesssim 10^{44}$ erg~s$^{-1}$ during the first six months after its UVOIR peak \citep{Renault-Tinacci2018.slsn.gamma}. However, these limits are not constraining on the expected leakage of the nebula energy in gamma-rays. At old ages, if the central engine exists, the X-ray signal will eventually emerge out due to the dilution effects, rather than the ionization breakout \citep{Margalit2018.cloudy}. Therefore, despite the predicted timescale $>$100 years, continued monitoring is essential. Besides the X-ray signal, we note that the radio signal is also a potential smoking gun \citep{Murase2016.pwn,Omand2018.radio.slsn}. Theoretical models or simulations to predict SEDs in various scenarios are necessary to distinguish the observed signals. The best candidates for future observations are the increasing number of very nearby events.

\acknowledgments

We thank G. Migliori for sharing expertise with XMM-SAS.
K.B. and R.C. acknowledge support from National Aeronautics and Space Administration (NASA) \textit{XMM-Newton} grant number 80NSSC18K0665.

\vspace{5mm}
\facility{XMM}

\software{BOXFIT \citep{vanEerten2012.afterglow.simulations}, MOSFiT \citep{Guillochon2017.mosfit}, SAS (https://www.cosmos.esa.int/web/xmm-newton/sas-news)
          }

\bibliographystyle{aasjournal.bst}
\bibliography{references.bib}

\end{document}